\documentclass[letter]{aa} 
\usepackage{graphicx}
\usepackage[utf8]{inputenc}
\usepackage{caption}
\usepackage{textcomp}
\usepackage{txfonts}
\usepackage{float}
\usepackage{natbib}
\bibpunct{(}{)}{;}{a}{}{,} 

\def\aap{A\&A}
\def\apjs{ApJS}

\newcommand{\tablenotea}[1]{\parbox{15.5cm}{\indent \footnotesize{#1}}}

\newcommand{\kms}{\mbox{km~s$^{-1}$}}

\newcommand{\s}{\mbox{$''$}}

\newcommand{\secp}{\mbox{\rlap{.}$''$}}

\newcommand{\irc}{IRC\,+10216}

\newcommand{\rasecp}{\mbox{\rlap{.}$^{\rm s}$}} 

\begin{document}

   \title{Clues to NaCN formation 
   \thanks{Based on observations carried out with 
   ALMA and the IRAM 30m Telescope. ALMA is a partnership 
   of ESO (representing its member states), NSF (USA) and NINS 
   (Japan), together with NRC (Canada) and NSC and ASIAA (Taiwan), 
   in cooperation with the Republic of Chile. The Joint ALMA Observatory 
   is operated by ESO, AUI/NRAO and NAOJ. IRAM is supported by INSU/CNRS (France), 
   MPG (Germany) and IGN (Spain). This paper makes use of the following 
   ALMA data: ADS/JAO.ALMA\#2013.1.00432.S \& ADS/JAO.ALMA\#2016.1.01217.S.}}


      \author{G. Quintana-Lacaci
          \inst{1}
          \and
          J. Cernicharo \inst{1}
           \and
                L. Velilla Prieto \inst{1}
           \and
          M. Ag\'undez \inst{1}
          \and
          A. Castro-Carrizo \inst{2}
           \and
         J.P. Fonfr\'ia \inst{1}  
         \and
         S. Massalkhi \inst{1} 
          \and
         J.R. Pardo \inst{1}  
          }
   \institute{Instituto de Ciencia de Materiales de Madrid, 
   Sor Juana In\'es de la Cruz, 3, Cantoblanco, 28049 Madrid, Spain.\\
              \email{g.quintana@icmm.csic.es} 
   \and 
             Institut de RadioAstronomie Millim\'etrique, 300 rue de la Piscine, 38406 Saint Martin d\'H\'eres, France\\
              }
   \date{Received September 15, 1996; accepted March 16, 1997}

 
  \abstract
   {ALMA is providing us essential information on where certain molecules form. Observing where these molecules emission arises from, the physical 
   conditions of the gas, and how this relates with the presence of other species allows us to understand the formation of many species, and
   to significantly improve our knowledge of the chemistry that
occurs in the space. }
   {We studied the molecular distribution of NaCN around \irc, a molecule detected previously, but whose origin is not clear. High angular resolution
   maps allow us to model the abundance distribution of this molecule and check suggested formation paths.}
   {We modeled the emission of NaCN 
   assuming local thermal equilibrium (LTE) conditions. These profiles were fitted to azimuthal averaged intensity profiles to obtain an abundance distribution of NaCN. }
   {We found that the presence of NaCN seems compatible with the presence of CN, probably as a result of the photodissociation of HCN,
   in the inner layers of the ejecta of \irc. However, similar
as for CH$_3$CN, current photochemical models fail to reproduce this CN reservoir.  
   We also found that the abundance peak of NaCN appears at a radius of $3 \times 10^{15}$cm, approximately where the abundance of NaCl, suggested to be the parent
   species, starts to decay. However, the abundance ratio shows that the NaCl abundance is lower than that obtained for NaCN. We expect that the LTE assumption might
   result in NaCN abundances higher than the real ones. Updated photochemical models, collisional rates, and reaction rates are essential to determine the
   possible paths of the NaCN formation.}
   {}


   \maketitle
%

\section{Introduction}

During its few cycles of operation, the ALMA interferometer has proved to be an incomparable tool
for the study of molecule formation in space.
A particular well-suited source for such studies is the C-rich AGB star IRC\,+10216, one of the closest 
evolved stars, located at a distance of $\sim$ 123\,pc from us \citep{IRC_distance}. While this object has been intensively studied, 
in particular in the field of the astrochemistry 
\citep[see, e.g.,][etc]{cernicharo2000,cerni2014b,innerlayer}, essential information on the exact location of the different
molecular reservoirs has remained inaccessible for years because
of the limited angular resolution of the instruments operating 
at (sub-)millimeter (mm) wavelengths. 

Only recently has the scientific exploitation of ALMA allowed us to understand with unprecedented detail the 
structure and kinematics of the circumstellar envelope of \irc\ \citep{cerni2014,Decin2015,NaCl_cycle0,Guelin2017} as well as 
the exact regions where the different species are formed
\citep{CH3CN,CarbonChains,SiSinprep}.

The location of the different molecules in the ejecta is crucial to understand the chemical processes that are at work in IRC\,+10216, in particular, and in circumstellar
envelopes in general.
In this work we study the distribution of the metal-bearing molecule NaCN around IRC\,+10216 and explore
its probable origin and the limitations of current chemical 
models. This work highlights once again the deep importance of the synergy between high-spatial resolution observations and
laboratory and theoretical work to obtain collisional rates and chemical reaction rate constants.

\begin{table*}
 \begin{center}
 \caption{Parameters of the observed lines.}
 \begin{tabular}{l l c c c c c c c}
 \hline\hline
  &&&&&\multicolumn{2}{c}{Low spatial resolution}& \multicolumn{2}{c}{High
spatial resolution}\\
  \hline
   Molecule &Trans. & Freq (MHz) & $E_\mathrm{up}$ (K) &S$_{ij}$ & Beam & P.A.($^\circ$) & Beam & P.A.($^\circ$) \\
  \hline
  NaCN & 6$_{1,6}$--5$_{1,5}$   &90394.38&17.6  & 5.83237 & 3\secp9$\times$2\secp7&89.9 &0\secp8$\times$0\secp6      &         39.2\\
  NaCN & 6$_{0,6}$--5$_{0,5}$ &93206.09&15.7    & 5.99388 & 4\secp3$\times$2\secp7&72.1 &0\secp6$\times$0\secp6      &         15.2\\
  NaCN & 6$_{2,4}$--5$_{2,3}$&94334.80&25.4     & 5.33293 &4\secp3$\times$2.6 &72.4 & 0\secp6$\times$0\secp6            &    40.4\\
  NaCN & 6$_{1,5}$--5$_{1,4}$&96959.81&18.7     & 5.83223 &3\secp9$\times$2\secp6 &92.7 &0\secp8$\times$0\secp7      &          20.0\\
  \hline 
  NaCN &  9$_{1,8}$--8$_{1,7}$ & 145075.57&37.3 & 8.88452 & 1\secp4$\times$1\secp1* &   52.4 &&\\
  NaCN &  10$_{2,8}$--9$_{2,7}$&  158616.77 &51.2 &9.59903 &1\secp3$\times$1\secp0* &   65.1 &&\\
  \hline 
  NaCl &$7-6$&  91169.88261&  17.5&  7.0& -- & -- & 0\secp8$\times$0\secp6     &           38.8\\
   \hline
   CH$_3$CN &$6_3-5_3$& 110364.35314& 82.8& 9.0& -- & -- & 0\secp8$\times$0\secp7 &                39.1\\
   CH$_3$CN &$6_0-5_0$& 110383.49871& 18.5& 6.0& -- & -- & 0\secp8$\times$0\secp7 &                38.9\\
   \hline
  
   \hline

  \label{Obs}
 \end{tabular}
 \tablenotea{\emph{High spatial resolution:} short-spacing and ALMA compact data merged. \emph{Low spatial resolution:} short-spacing
and ALMA compact and extended data merged. *: only ALMA visibilities.}
\end{center}
\end{table*}


\section{Observations and NaCN distribution}

We carried out a $\lambda~3$~mm spectral survey of IRC\,+10216 
with ALMA band\,3 during Cycle\,2, covering the 
frequency range 84.0-115.5 GHz.
In addition, we recently obtained a mosaic with ALMA covering
selected frequencies at 2\,mm. 
These observation are described in 
detail in Appendix\,\ref{sect:obs}.
 
We obtained interferometric maps of the NaCN transitions 
presented in Table\,1, both merging ALMA compact configuration data
and the on-the-fly (OTF) maps obtained with the IRAM 30m telescope, and 
the ALMA data in the compact and extended configurations with 
the OTF single-dish maps for the observations of Cycle\,2.
This resulted in two sets of maps, one with a typical angular resolution of 4$''\times 3''$ , and 
another with a higher angular resolution of $0\secp8\times0\secp7$.

The sensitivity of the individual maps of each transition is not high enough 
to have a clear view of the distribution of the NaCN emitting 
gas. The brightness distribution that is visible in the velocity channels suggests a spherical hollow shell-like distribution. 
In order to increase the signal-to-noise ratio (S/N) and to confirm this brightness distribution, 
we stacked the emission from the NaCN lines presented in Table.\,\ref{Obs}.
In the case of the low angular resolution maps, this stacking already allowed 
us to confirm the hollow shell gas distribution suggested by the single maps
(see Fig.\ref{map}). However, in the case of the high spatial resolution maps, 
the stacking alone was not enough. 
We therefore smoothed the spectral resolution to 5\kms to increase the S/N.
This high spatial -- low spectral resolution map is presented in
Fig.\,\ref{mapsce}.

\begin{figure}[th!]
   \centering
   \includegraphics[angle=0,width=9cm]{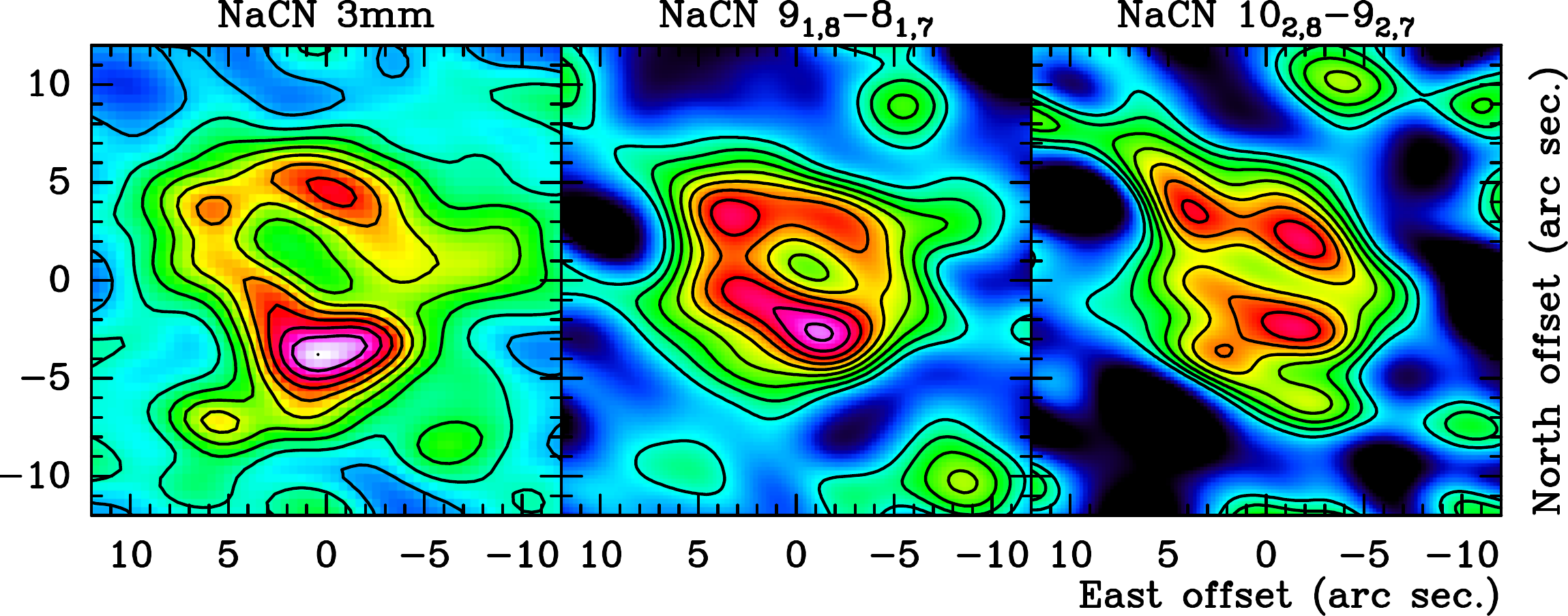}
   \caption[]{Comparison of the central channel of the low spatial
resolution 3mm stacked map with maps of the ALMA Cycle\,4 observations at 2\,mm.
   The lowest contour corresponds to 10\% of the peak flux, and the rest of contours are equally spaced in jumps of 10\% 
   with respect to the first contour. For absolute intensities see Figs.\ref{map},\ref{map2mm1}, and \ref{map2mm11}. }
              \label{2mm}%
\end{figure}

\begin{figure}[h]
   \centering
   \includegraphics[angle=0,width=8cm]{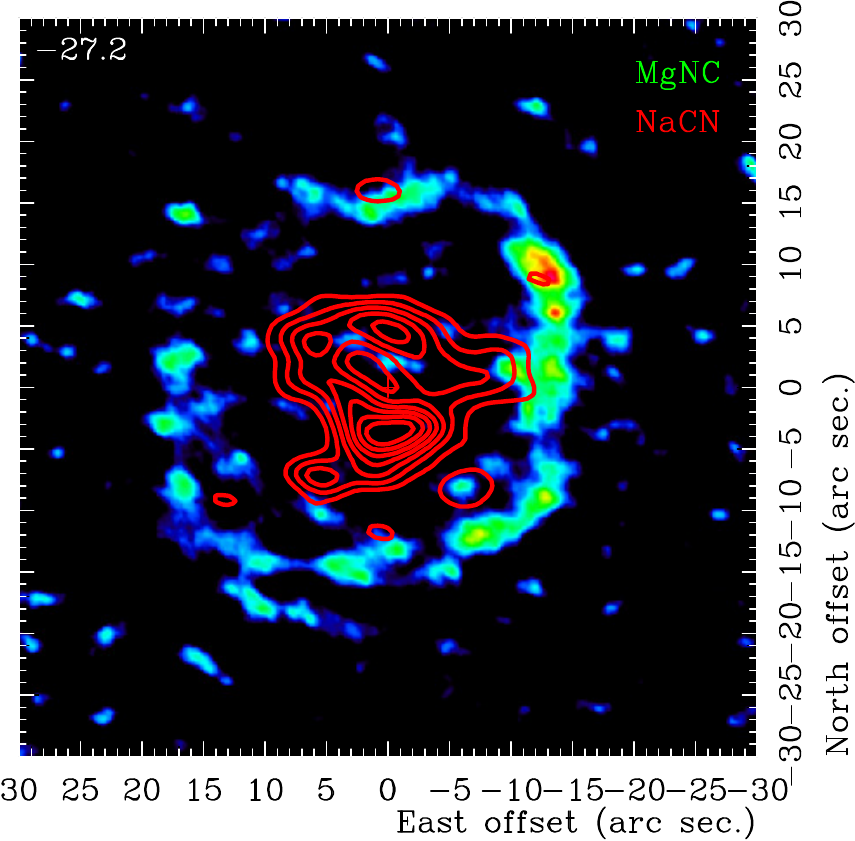}
   \caption[]{Comparison of the extent of emission between NaCN (red contours, see Fig\,\ref{map}) and MgNC (color scale, see Fig.\ref{MgNC}). }
              \label{MgNC_comp}%
\end{figure}

In addition, we merged the visibilities of the seven fields for Cycle\,4 NaCN 
data and obtained cleaned maps with an angular 
1\s. 
These maps presented negative fluxes and a smaller extent than the 3\,mm NaCN 
data, confirming the flux loss. Similar as for the 3\,mm data, we reduced the spatial 
resolution to increase the S/N. In particular, we imposed in the cleaning algorithm
the synthezised beam to obtain the same 
spatial resolution as in the map presented in Fig.\,\ref{map}. These 
maps are presented in Figs.\,\ref{map2mm1}\, and\,\ref{map2mm11}.
The NaCN brightness distribution of the stacked map of the 3\,mm lines and from the 
two NaCN transitions observed at 2\,mm show the very same structure, confirming 
the presence of the hollow shell mentioned above
(Fig.\,\ref{2mm}).




\section{NaCN modeling}


In order to fit the observed gas distribution, we assumed the density and temperature profiles
adopted by \citet{NaCl_cycle0}, { slightly modified to follow the temperature law deduced by \citet{Guelin2017}}. 
Similarly to what was done in that work, the abundance profile was adjusted 
to the observation using as starting point the abundance profiles derived by \citet{innerlayer}.
We only aimed at fitting the 3\,mm data because for these transitions all the flux has been recovered by merging
the ALMA visibilities with the short-spacing data.

Since there are no available collisional rates for NaCN, we used a local thermodynamical equilibrium (LTE) multi-shell 
approach carried out with the MADEX code \citep{MADEX} to reproduce the observations. 
In this approach, we solve the level populations at the different radii, assuming LTE conditions. 
When this is solved, the synthetic profiles are obtained 
by solving the ray tracing and convolving with the beam of the telescope.

\begin{figure}[th!]
   \centering
   \includegraphics[angle=0,width=7cm]{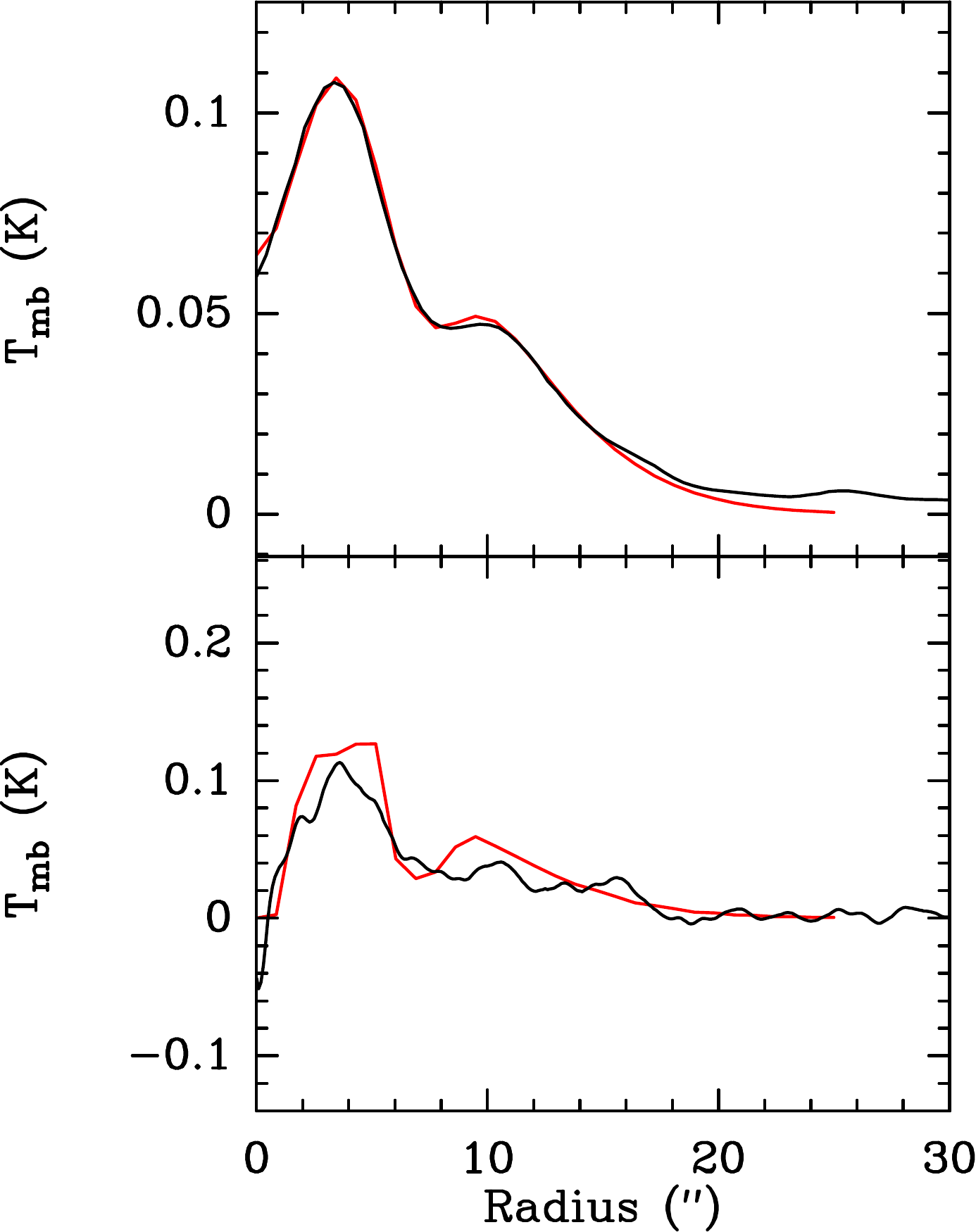}
   \caption[]{\emph{Top:} Model fitting (red line) of the azimuthally averaged emission of the stacked NaCN lines presented in Table
\ref{Obs} for the low spatial resolution maps
   (Fig.\,\ref{map}). \emph{Bottom:} Model fitting (red line) of the azimuthally averaged emission of the stacked NaCN lines presented in Table \ref{Obs} for the high spatial resolution maps
   (Fig.\,\ref{mapsce}).
   }
              \label{azave}%
\end{figure}

As for fitting the abundance profile, we assumed that the emission distribution is mainly spherical. 
Therefore, we obtained the azimuthal averaged emission of the NaCN stacked map for both
the high- and the low-resolution maps. We aimed at fitting both emission profiles at the same time, taking into account
that the S/N difference of the low-resolution and high-resolution maps is significant. A good fitting of the high S/N emission
profile was mandatory, while a reasonable fit of the low S/N profile was enough. These fits are shown in Fig.\ref{azave}.

\section{Discussion}

The location of the different metal-bearing cyanides can be separated into two main groups. 
Some of them, such as MgNC \citep{guelin1993} and HMgNC \citep{HMgNC}, are found to appear far from the
star's photosphere, forming a shell-like distribution with a radius of $\sim$ 15$''$. This structure 
has been confirmed in MgNC by our present ALMA observations (see Fig.\ref{MgNC}). In contrast, species such as
NaCN were found to present compact emission when observed with a
spatial resolution of $\sim$3${''}$ \citep[][]{guelin1993,guelin1997}. 
These latter results suggested that the formation of NaCN takes place in 
chemical equilibrium conditions in the vicinity of the stellar photosphere.
However, the maps we present here reveal a small inner hole with a radius $\sim$ 1.5$''$ in the brightness distribution of 
NaCN, suggesting that it is formed in regions where 
the gas has left the chemical equilibrium regime.

These two-fold distribution of the metal cyanides has been theoretically explained by \citet{Petrie1996}
based on the physical characteristics of the parent species that combine with 
CN. In particular, metal-bearing species such as AlCl, NaCl, or KCl are closed-shell
molecules that form in the inner hot regions of the envelope, while other species
containing Mg, Fe, or Si are open-shell radicals that could react with other neutral molecules
in regions with low temperatures. Reactions involving CN and the former group of species 
would therefore result in metal cyanides in the innermost regions of the ejecta, while those involving the latter 
group will result in cyanides forming extended shell-like structures like those cited above.
A comparison of the extent of these diferent cyanides is shown in Fig.\ref{MgNC_comp}.


In the particular case of NaCN, \citet{Petrie1996} suggested the following pathway:

\begin{equation}
 \mathrm{NaCl + CN \rightarrow NaCN + Cl} 
 .\end{equation}

\begin{figure}[th!]
   \centering
   \includegraphics[angle=0,width=7.5cm]{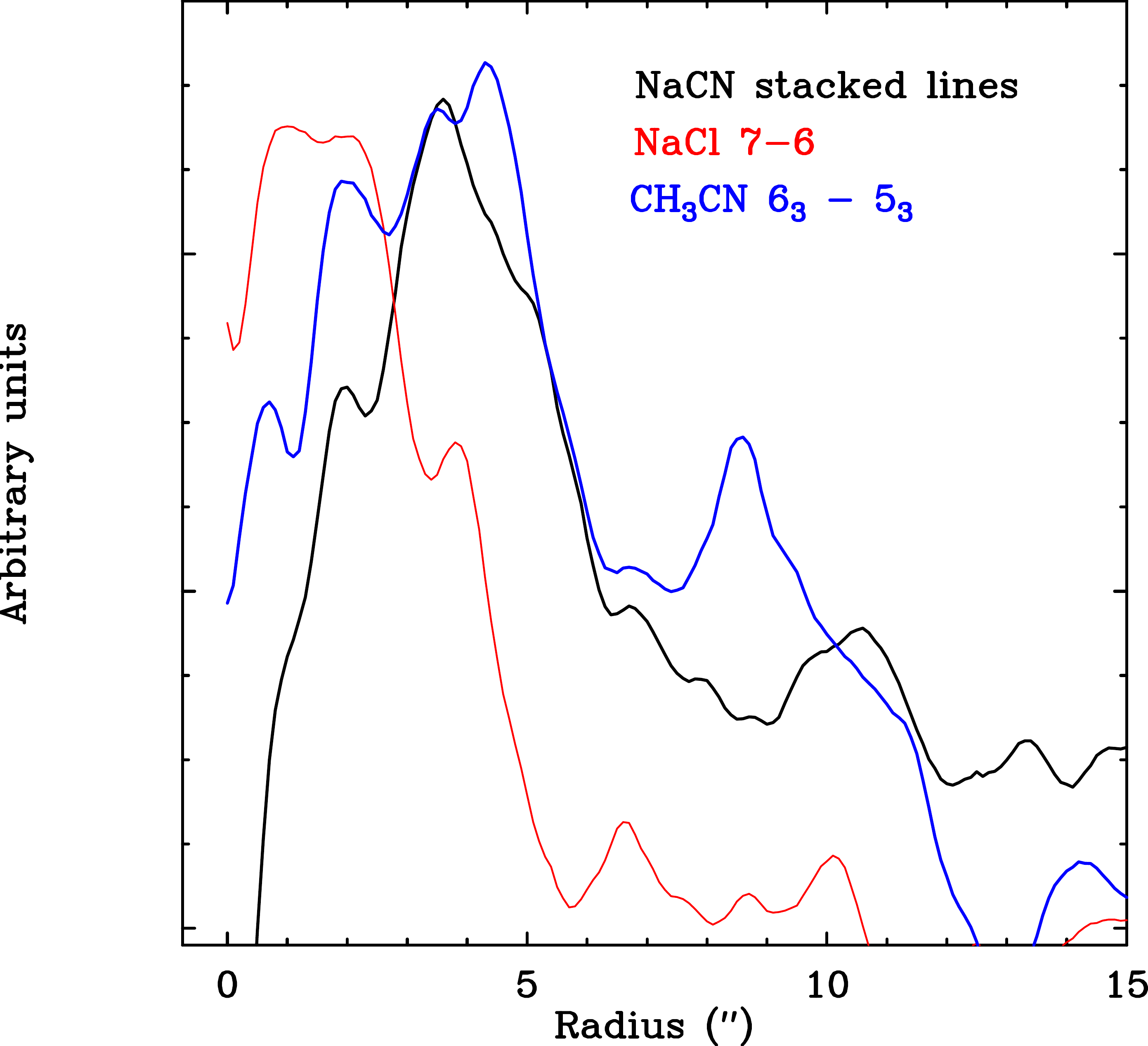}
   \caption[]{Comparison of the azimuthally averaged intensity profiles of the NaCN maps (black line), 
   the NaCl 7-6 emission map (red line), and the CH$_3$CN $6_3-5_3$ (blue line). }
              \label{comp}%
\end{figure}

Since NaCl is a closed-shell species that is abundant in the innermost regions of the envelope \citep{NaCl_cycle0}, 
this reaction could take place when a significant amount of CN is available.
The distribution of the NaCl $J=7-6$ emission, with
an energy of the upper level ($E_\mathrm{up}$) similar to the NaCN transitions 
we analyzed here and obtained within the same ALMA Cycle\,2 project 
(see Appendix\,\ref{sect:obs} and Table\,\ref{Obs} for details), 
showed that its brightness distribution is complementary to that of NaCN, that is, when NaCl emission fades, NaCN emission rises. 
This is shown in Fig.\,\ref{comp} by comparing the azimuth-averaged emission of the central velocity channel for the different 
transitions. 
This seems to support the assumption that NaCl is a parent molecule for the formation of NaCN.

On the other hand, the models suggest that CN is present only near the photosphere 
and in the outer layers of the CSE as a result of the photodissociation of
HCN \citep{Lucas1995}. Recent ALMA CN maps confirm that the CN emission appears at typical radii of
$\sim 15''$  \citep{CarbonChains}.

Recently, \citet{CH3CN} showed that the CH$_3$CN spatial distribution was
unexpectedly located in an inner hollow shell. 
To study the possible relation of the formation of NaCN with that of CH$_3$CN,  we
have compared the brightness distribution
of this molecule 
with that of the stacked NaCN map.
While CH$_3$CN 6$_0$-5$_0$ has a excitation temperature similar to that of the NaCN transitions we present here, 
it is blended with the 6$_1$-5$_1$. We therefore used the unblended transition 6$_3$-5$_3$ for the comparison.
The two azimuth-averaged maps present very similar distributions, suggesting 
that the parent molecule responsible for both molecules might be similar. In particular, an
injection of CN at a radii of $\sim 1\secp5-2''$ could explain the distribution of both CH$_3$CN and NaCN. 
Such an injection would also affect other species, in particular, HC$_3$N. However, since the rate constant of reaction (1) is
unknown, we cannot estimate the balance between the different reactions involving CN. 

Furthermore, \citet{CH3CN} showed that the model presented by \citet{watermarce}, which takes into account the effect of penetrating
UV photons in the innermost layers of the circumstellar envelope
(CSE) around \irc\ fails to explain the CH$_3$CN distribution. Therefore, to understand
the CN source and reaction distribution, a detailed new model is mandatory. This model will be developed in a forthcoming paper.

The results shown in Fig.\,\ref{comp} clearly suggest that the reaction proposed by \citet{Petrie1996} is plausible and that
a certain amount of cyanide is freed by photodissociation of HCN or other CN-bearing species
with weaker bounds than NCCN, not detected in space so far, and that it reacts rapidly to form NaCN and CH$_3$CN.

\section{Results}

\begin{figure}[th!]
   \centering
   \includegraphics[angle=0,width=8cm]{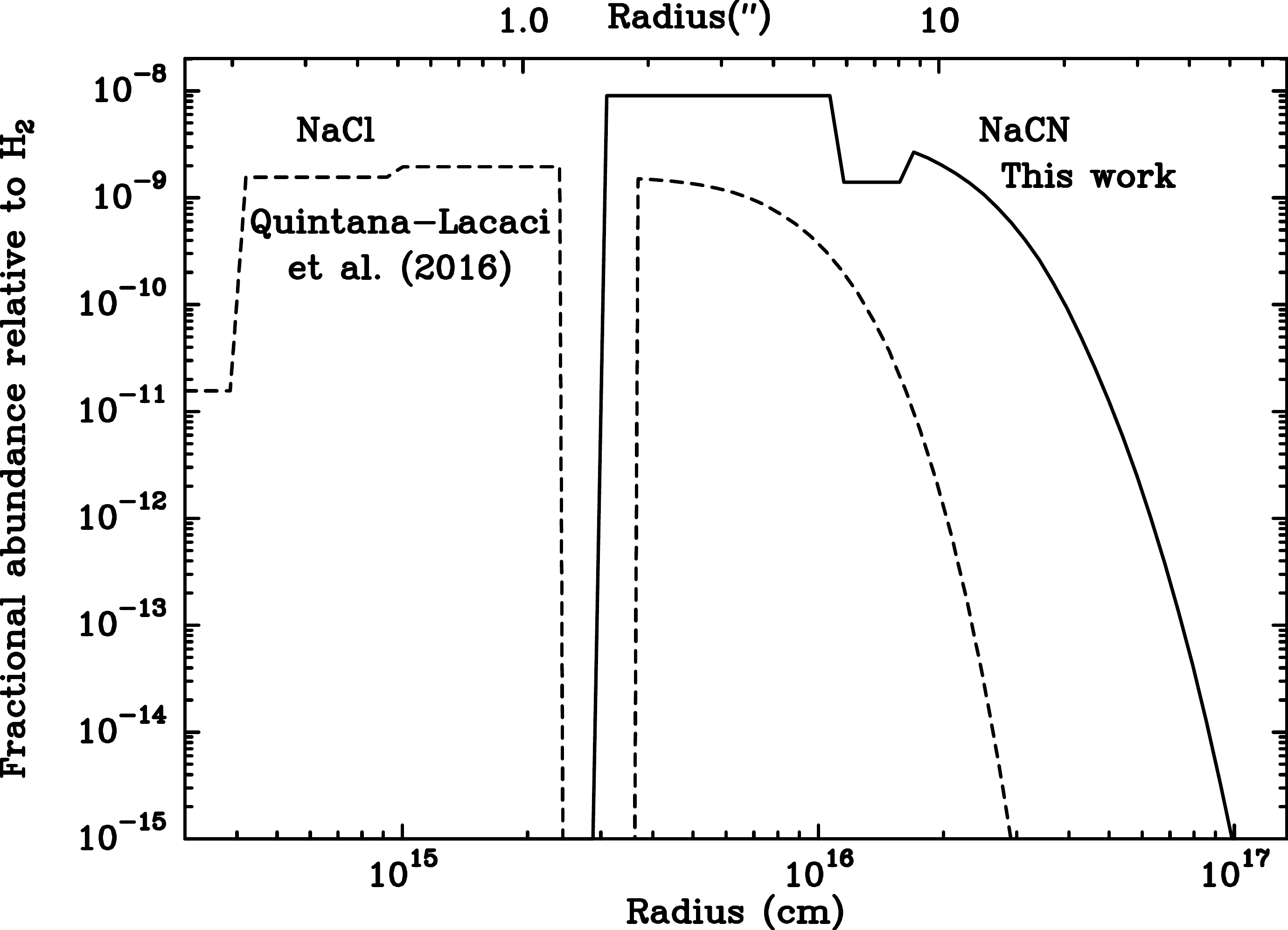}
   \caption[]{\emph{Solid line:} NaCN abundance profile obtained assuming LTE conditions. \emph{Dashed line:} NaCl abundance profile
   obtained by \citet{NaCl_cycle0}.}
              \label{abun}%
\end{figure}

The abundance profile obtained for NaCN is presented in Fig.\,\ref{abun}
As suggested by the NaCN emission maps (see Fig.\,\ref{map} and \ref{mapsce}), we found that NaCN
arises at a radius of $3\times10^{15}$\,cm.

To check the abundance relation between NaCN and NaCl, we compared the abundance profile obtained here with that
derived by \citet{NaCl_cycle0} for NaCl (dashed line in Fig.\,\ref{abun}).
This comparison shows that 
the tentative precursor has a lower abundance than the
resulting species. This might indicate that reaction (1) is not the main formation path of NaCN. 
However, we have to keep in mind that while the NaCl abundance profile was accurately derived by solving the level population in non-LTE conditions, 
that of NaCN has been derived assuming LTE.

At the regions where NaCN abundance rises ($3\times10^{15}$\,cm) $T_{\mathrm{K}} \sim 160$\,K \citep{Guelin2017}.
This means that because LTE assumes $T_\mathrm{ex}=T_\mathrm{K}$,
the high-excitation lines are favored over low-excitation lines
such as we studied here. The regions where, in LTE, these low-excitation transitions are
expected to dominate lie at radii $\sim 2.5 \times 10^{16}$cm. However, as shown by \citet{innerlayer}, metal-bearing species studied by these authors leave the LTE regime
at the regions where NaCN emission arises. 
{ Furthermore, these authors showed that 
the line intensity ratios from the NaCN lines observed are not compatible with an LTE regime.}
Therefore, we might expect non-LTE 
modeling to derive { lower }values of $T_\mathrm{ex}$ and therefore higher level populations, 
higher intensities, and lower abundances for the transitions observed.
This lower abundance would then conciliate the NaCl and NaCN abundance ratio confirming reaction (1) as the main formation path for NaCN.

{ Another factor that might affect the estimate of the NaCN abundance is the effect of 
the IR pumping on the NaCN excitation. This effect has been found to be important for other species
such as NaCl \citep{NaCl_cycle0}. However, there is no information available in the literature about 
the IR ro-vibrational spectrum of NaCN that might help to estimate the effect of the IR pumping.}


\section{Conclusions}

We have obtained interferometric maps with high and intermediate angular resolution of the metal-bearing molecule
NaCN. As shown by \citet{guelin1997} and \citet{Petrie1996}, this molecule emission arises in the inner regions of the envelope of \irc.
New maps have shown that this emission presents an inner hole that has previously not been detected.
Furthermore, emission from NaCl, as well as its abundance, suggest that when NaCl declines, NaCN rises. 
However, two problems prevent us from confirming reaction (1) as the main formation path of NaCN. 

First, the origin of CN is not clear. \citet{CH3CN} did not succeed to model the abundance of this species taking into account
the penetration of UV photons into inner layers as a source of CN. A new photochemical model is required to simultaneously explain the 
source of the CN  and its impact on CH$_3$CN, NaCN, and HC$_3$N formation. Obtaining a reaction 
rate for (1) is essential to solve the competition for CN for the different chemical paths. 

Second, the derived abundance of NaCN seems to be higher than that of NaCl. This might be a sign of different 
parent species, or, more probably, an artifact derived from the LTE assumption. Obtaining collisional rates for NaCN would allow us to 
solve this problem.

\begin{acknowledgements}
      
The research leading to these results has received funding from the European Research Council
under the European Union's Seventh Framework Programme (FP/2007-2013) / ERC Grant
Agreement n. 610256 (NANOCOSMOS). 
We would also like to thank the Spanish MINECO for funding support from grants CSD2009-00038, 
AYA2012-32032 \& AYA2016-75066-C2-1-P. M.A. also ackowledges funding support from the Ramón y Cajal programme of Spanish MINECO (RyC-2014-16277). 
\end{acknowledgements}

\bibliographystyle{aa} 
\bibliography{new} 

\begin{appendix}

\section{Observations}
\label{sect:obs}
The observations were obtained with compact and extended 
array configurations, with baselines in the range $\sim 12 - 300$\,m and $\sim 30 - 2000$\,m, respectively.
 The field of view (FoV) of the 12m ALMA 
antennas ranges from $\sim$69$''$ at 84 GHz to $\sim$50$''$ at 115.5 GHz. 
Additional observations were performed with the IRAM 30m telescope to 
recover the flux filtered out by the interferometer. Observations were 
centered on the position of the star, with coordinates J2000.0 
R.A.=09$^{\rm h}$47$^{\rm m}$57\rasecp446 and Dec.=13$^{\circ}$16$'$:43\secp86, 
according to the position of the $\lambda$~1~mm continuum emission peak 
\citep{cerni2014b}. A detailed description of the spectral survey 
will be presented elsewhere (Cernicharo et al., in prep). 
The data were calibrated using the CASA\footnote{http://casa.nrao.edu/} software package, and
imaged and cleaned with GILDAS\footnote{http://www.iram.fr/IRAMFR/GILDAS/} software package. We used the SDI cleaning algorithm
since HOGBOM could generate artificially clumpy structures for high 
spatial resolution observations.

\begin{figure*}[h!]
   \centering
   \includegraphics[angle=0,width=16cm]{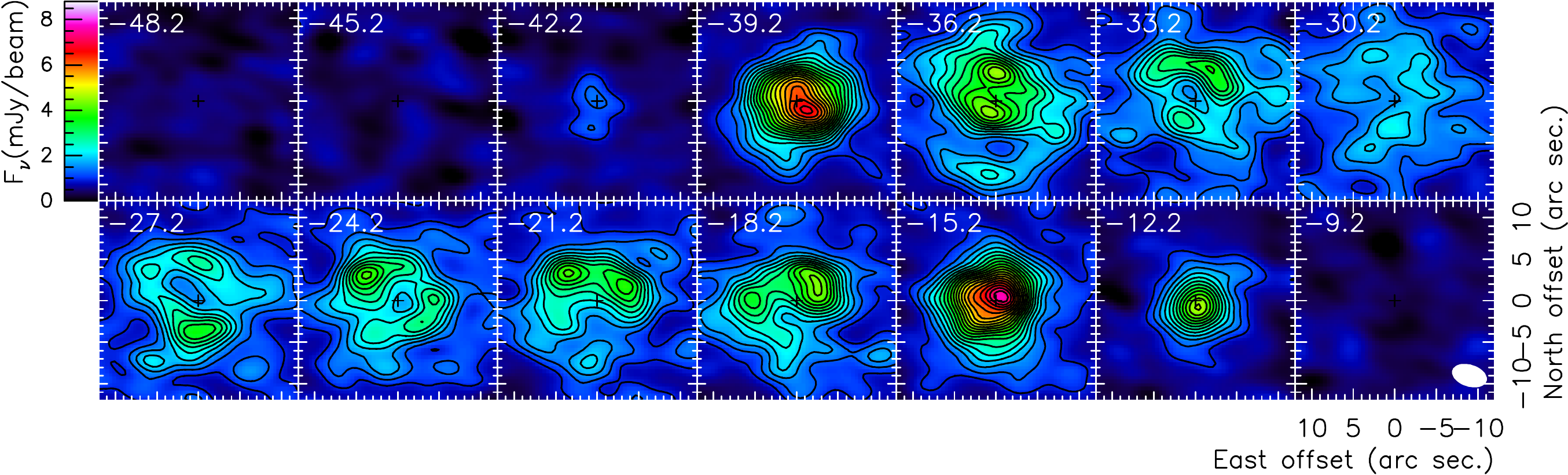}
   \caption[]{Low spatial resolution interferometric map of the NaCN stacked emission of the transitions at $\lambda~3$~mm presented in Table\,1. 
   In the upper left corner of each panel we note the v$_{LSR}$ of the channel ($V_\mathrm{sys}=-26.5\kms$). 
   The lowest contour corresponds to a value of 3$\sigma,$ and the rest of the contours are equally spaced in jumps of 2$\sigma$ 
   with respect to the first contour. The rms of the map is $\sigma= 0.8 $\,mJy\,beam$^{-1}$. The beam size is drawn in the last panel. 
   The HPBW is $4\secp3\times 2\secp6$ with a P.A. of 72$^\circ$. 
 }
              \label{map}%
\end{figure*}

\begin{figure*}[h]
   \centering
   \includegraphics[angle=0,width=18cm]{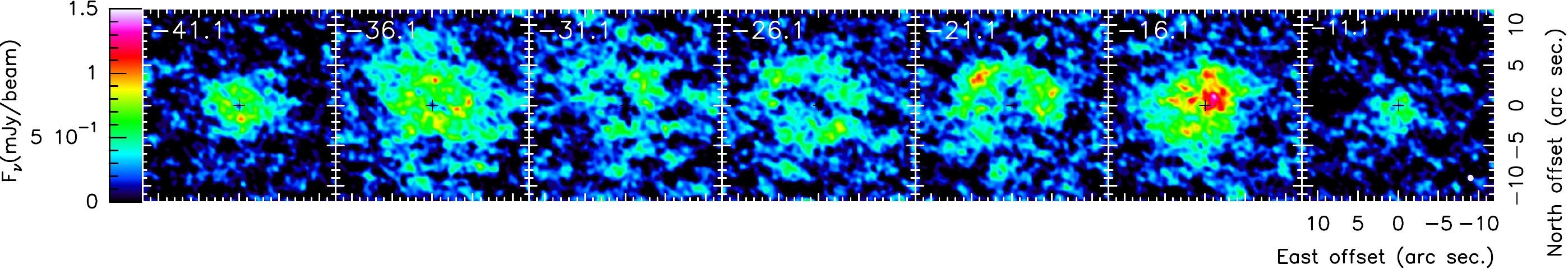}
   \caption[]{High spatial resolution interferometric map of the NaCN stacked emission of the transitions presented in Table\,1. 
   In the upper left corner of each panel we note the v$_{LSR}$ of the channel ($V_\mathrm{sys}=-26.5\kms$). 
   The rms of the map is $\sigma= 0.76 $\,mJy\,beam$^{-1}$. The beam size is drawn in the last panel.  The HPBW is $0\secp8\times 0\secp7$ with a P.A. of 20$^\circ$.
   The flux density scale is in    Jy\,beam$^{-1}$.   }
              \label{mapsce}%
\end{figure*}

For the emission 
lines studied here, data from the ALMA compact and extended configurations 
were merged, after continuum 
subtraction, with the short-spacing data obtained with the IRAM 30m telescope.
In particular, since the NaCN emission is relatively weak, we present low spatial
resolution NaCN maps with high S/N, obtained by merging the ALMA compact configuration and 
short-spacing data, and high spatial resolution maps with low S/N, obtained when merging both 
ALMA configurations plus the short-spacing data (see Table\,1 for details).
In order to increase the S/N, we stacked the NaCN emission of the different lines lying at 3\,mm.
These maps are presented in Figs.\ref{map} and \ref{mapsce}.

\begin{figure*}[h]
   \centering
   \includegraphics[angle=0,width=18cm]{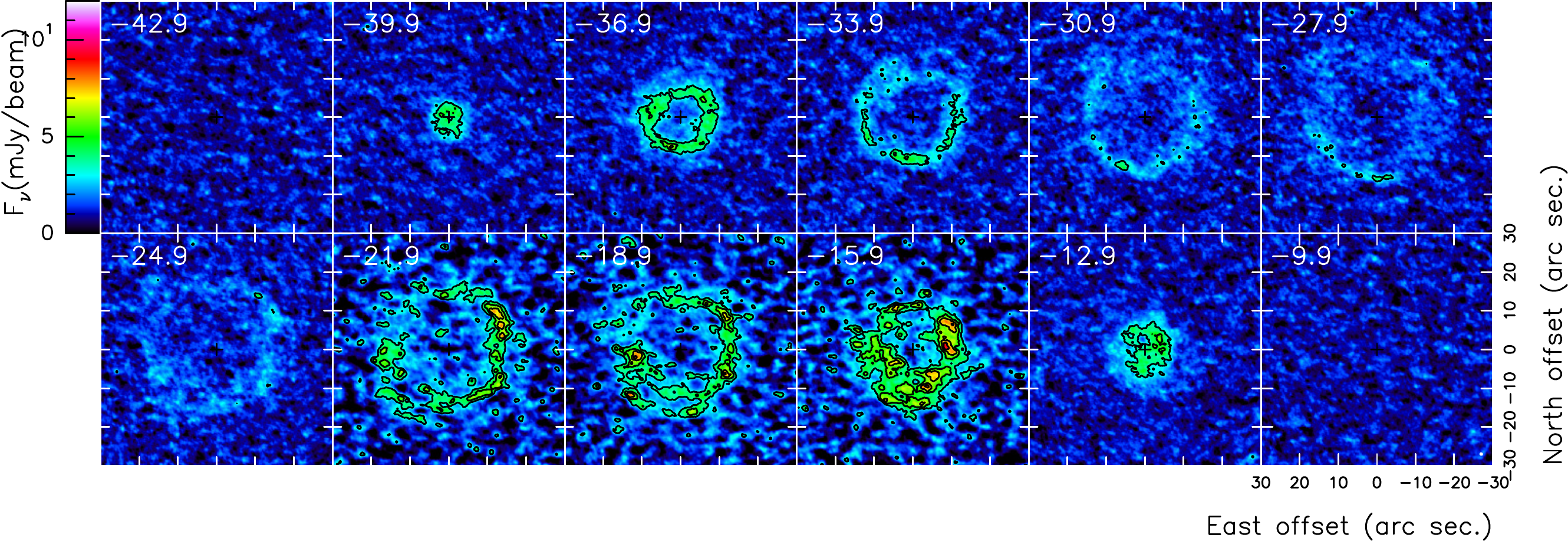}
   \caption[]{Interferometric map of the MgNC $8_{17/2}-7_{15/2}$ emission of the transitions observed within ALMA project 2013.1.00432.S. 
   In the upper left corner of each panel we note the v$_{LSR}$ of the channel ($V_\mathrm{sys}=-26.5\kms$). 
   The lowest contour corresponds to a value of 3$\sigma,$ and the rest of contours are equally spaced in jumps of 2$\sigma$ 
   with respect to the first contour.
   The rms of the map is $\sigma= 0.9$\,mJy\,beam$^{-1}$. The beam size is downgraded to that of Fig.\,\ref{map} and is drawn in the last panel.  The HPBW is $0\secp9\times 0\secp8$ with a P.A. of 33$^\circ$.
   The flux density scale is in mJy\,beam$^{-1}$.   }
              \label{MgNC}%
\end{figure*}

We also present a map of MgNC $8_{17/2}-7_{15/2}$, observed at
the same time as the 3\,mm ALMA line survey.
As for NaCN emission map, we merged single-dish OTF data with those of the extended and compact ALMA configurations. 
This map is presented in Fig.\,\ref{MgNC}

\begin{figure*}[h]
   \centering
   \includegraphics[angle=0,width=18cm]{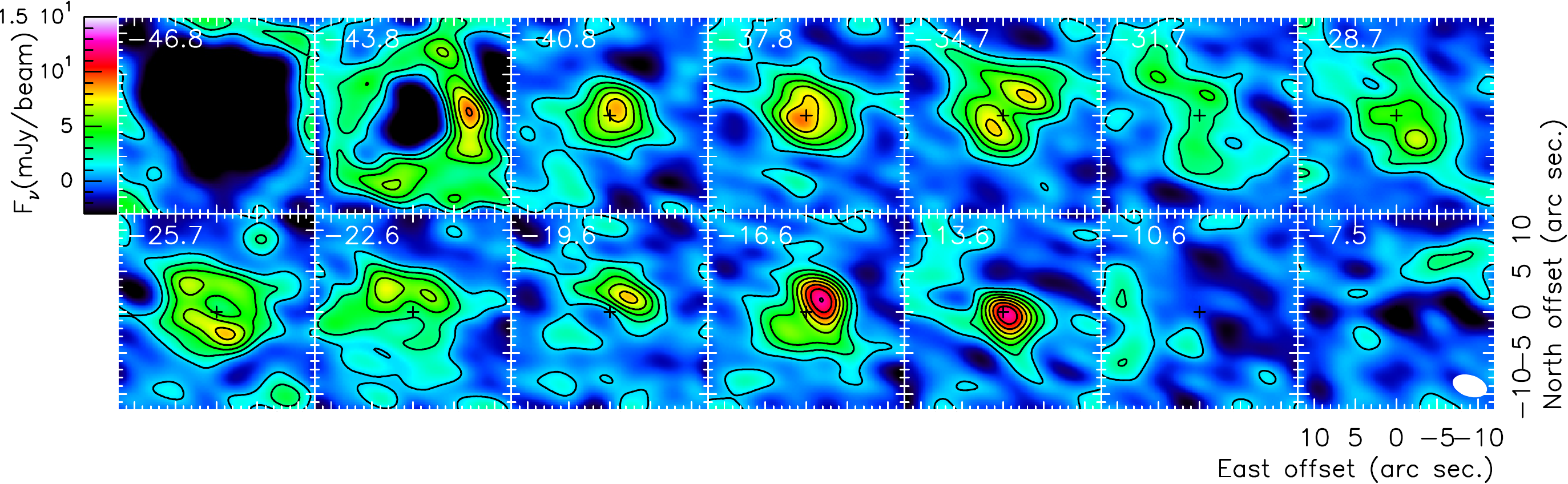}
   \caption[]{$\lambda$~2\,mm interferometric map of the NaCN 9$_{1,8}$--8$_{1,7}$ emission. 
   The emission of the line C$_3$H$_2$ $3_{1,2}-2_{2,1}$ at 145089.61104 MHz is partially blended with
   that of NaCN and can be seen in the first channels. The central channels of the NaCN are, however, free of 
   this pollution.
   In the upper left corner of each panel we note the v$_{LSR}$ of the channel ($V_\mathrm{sys}=-26.5\kms$). 
   The lowest contour corresponds to a value of 1 $\sigma,$ and the rest of contours are equally spaced in jumps of 1.5$\sigma$ 
   with respect to the first contour.
   The rms of the map is $\sigma= 1.1$\,mJy\,beam$^{-1}$. The beam size is downgraded to that of Fig.\,\ref{map} and is drawn in the last panel.  The HPBW is $0\secp8\times 0\secp7$ with a P.A. of 20$^\circ$.
   The flux density scale is in    Jy\,beam$^{-1}$.   }
              \label{map2mm1}%
\end{figure*}

\begin{figure*}[h]
   \centering
   \includegraphics[angle=0,width=18cm]{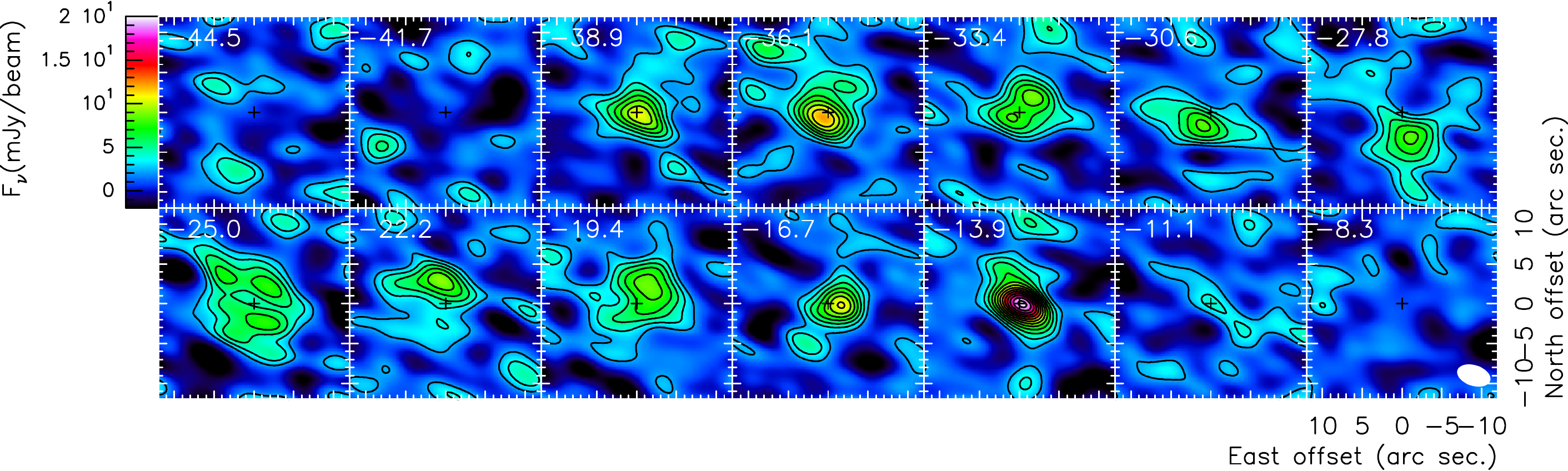}
   \caption[]{$\lambda$~2\,mm interferometric map of the NaCN 10$_{2,8}-9_{2,7}$ emission of the transitions presented in Table\,1. 
   In the upper left corner of each panel we note the v$_{LSR}$ of the channel ($V_\mathrm{sys}=-26.5\kms$). 
   The lowest contour corresponds to a value of 1$\sigma$ and the rest of contours are equally spaced in jumps of 1$\sigma$ 
   with respect to the first contour.
   The rms of the map is $\sigma= 1.5$\,mJy\,beam$^{-1}$. The beam size is downgraded to that of Fig.\,\ref{map} and is drawn in the last panel.  The HPBW is $0\secp8\times 0\secp7$ with a P.A. of 20$^\circ$.
   The flux density scale is in    Jy\,beam$^{-1}$.   }
              \label{map2mm11}%
\end{figure*}

In addition, we recently obtained ALMA Cycle-4 data that cover two
NaCN transitions at $\lambda~2$~mm  
(see Table\,1).
These observations consisted in a mosaic of seven
fields covering a FOV of $\sim$214\s. 
We note that for these transitions we did not have short-spacing data, and thus
the maps only rely on ALMA visibilities. Although
the NaCN emission is relatively compact, some flux is expected to be  
filtered out. The baselines of the interferometer were in the range $\sim 12 - 408$\,m.
For a detailed description of these ALMA-Cycle\,4 observations see Velilla Prieto et al. (in prep). 
These maps are shown in Figs.\,\ref{map2mm1} and \ref{map2mm11}.



\end{appendix}

\end{document}